\documentclass[twocolumn,showpacs,preprintnumbers,amsmath,amssymb,aps,prl]{revtex4}
\usepackage{graphicx}
\begin{document}

\title{Dynamical Ordering and Directional Locking For Particles 
Moving Over Quasicrystalline Substrates}   
\author{C. Reichhardt and C. J. Olson Reichhardt} 
\affiliation{
Theoretical Division,
Los Alamos National Laboratory, Los Alamos, New Mexico 87545, USA } 

\date{\today}
\begin{abstract}
We use molecular dynamics simulations to study the 
driven phases of particles such as vortices or colloids
moving over a decagonal quasiperiodic substrate. 
In the regime where the pinned states have quasicrystalline ordering, 
the driven phases can order into moving square or  
smectic states, 
or into states with aligned rows of both square and triangular
tiling which we term dynamically induced Archimedean-like tiling. 
We show that when the angle of the drive is varied with respect to the
substrate, 
directional locking effects
occur where the particle motion locks to certain angles. 
It is at these locking angles that the dynamically induced Archimedean
tiling appears. 
We also demonstrate that the different dynamical 
orderings and locking phases show pronounced changes
as a function of filling fraction.    
\end{abstract}
\maketitle

\vskip2pc
Quasicrystals have received intense study since their discovery due to
their unusual property of combining nonperiodicity with long-range
order \cite{SR}. 
Recently there has been growing interest in understanding
how interacting particles such as
vortices in type-II superconductors 
\cite{MiskoKemler,Sihr,Montero}
or charged colloidal particles \cite{Bechinger,Stark,M,C,G} 
order in the presence of a quasicrystalline substrate
created with nanolithographic or optical techniques. 
On a fully periodic substrate, these types of particles form
states that are commensurate with the periodicity of the substrate at 
fillings which meet integer or fractional matching conditions
\cite{Baret,Reichhardt,Harada,Olson2Frey}. 
In the case of vortices, the commensurate fillings are associated with
peaks in the critical current or the external force 
required to depin the vortices from the substrate \cite{Baret,Reichhardt}.
When vortices interact with quasicrystalline Penrose or decagonal pinning site
arrays,
a new set of peaks in the critical current appear at non-rational
fields in addition to the commensurate fields 
due to a novel type of ordering of the vortices on these 
substrates \cite{MiskoKemler,Sihr}. 
For colloids interacting with decagonal substrates, 
experiments show that strong substrates produce quasicrystalline order
of the colloids but weak substrates permit the triangular lattice favored
by the colloid-colloid interactions to form.
Remarkably, for intermediate substrate strengths 
a new type of ordering arises consisting of a combination
of triangular and square order.  This has been
termed an Archimedean-like tiling \cite{Bechinger,C}.   

Numerous experimental and simulation studies have been performed on the
dynamic phases of a collection of particles driven over a periodic
substrate.
One of the most striking observations is directional locking 
in which the particles cease to follow the direction of the drive and instead
lock to a symmetry direction of the substrate.
The locking effect produces
a series of pronounced steps in the velocity versus drive angle curve
\cite{ReichhardtA,Ahn}.
Directional locking effects were experimentally observed for colloids moving 
through a periodic optical trap array \cite{Korda,Lacasta} 
and for vortices driven at differing angles 
with respect to a pinning substrate \cite{G2}. For  
colloidal systems, directional locking can be used for practical
applications such as the fractionation of
different species of particles, only one of which locks to the symmetry
direction of the substrate
\cite{Lacasta,Drazer}. 
An open question is whether
directional locking can also occur for particles 
moving over quasiperiodic substrates.
It has already been shown that the nonequilibrium, driven state for
particles driven over random substrates can show dynamically induced
ordered states which are inaccessible to the static system.
For example, on strong random substrates which produce a disordered
pinned state, the driven state can partially 
order to a moving smectic or to a triangular lattice \cite{Koshelev,Giamarchi}.

Here we show that directional locking and dynamical ordering
can occur for particles moving over quasicrystalline substrates. 
The dynamical ordering occurs only for certain driving 
directions, and changes the system from 
a quasicrystalline pinned state
to a moving state with novel dynamical ordering that is similar to
a square and triangular tiling. 
We call this ordering a dynamically induced Archimedean tiling. 
Other types of ordering to moving smectic, moving square,
and moving liquid states are also possible as a function of 
filling and drive direction.
We specifically demonstrate the locking and ordering
for vortices in type-II superconductors, and note that the
same results appear for charged colloids moving 
over decagonal substrates \cite{Sn}.  

\begin{figure}
\includegraphics[width=3.5in]{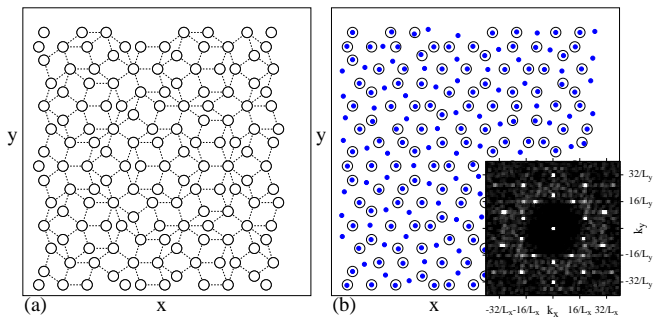}
\caption{
(a) The locations of the Penrose tiled pinning sites 
(open circles) for a portion of the sample.  Dashed lines indicate the
positions of the Penrose tiles used to define the pinning locations.
(b) The location of the vortices (filled circles) and
pinning sites (open circles) at 
$B/B_\phi=1.61$
in the same sample
at zero applied drive.  Here
$B_{\phi}$ is the field at which the number of vortices
equals the number of pinning sites.
Inset: The structure factor $S(k)$ of the vortex positions from (b) has
tenfold peaks indicative of quasicrystalline order. 
}
\end{figure}

We model a two-dimensional system of 
$N_v$ interacting particles in the presence of a pinning array with 
5-fold Penrose ordering,
as shown in Fig.~1(a).  
In the superconducting system we consider a sample 
of size 
$24\lambda \times\ 24\lambda$
with periodic boundary conditions, 
where $\lambda$ is the London penetration depth. 
The vortex motion evolves according to the following overdamped equation: 
$\eta\frac{d{\bf R}_i}{dt} = {\bf F}^{vv}_{i} + {\bf F}^{p}_{i} + {\bf F}^{ext}_{i} $. 
Here ${\bf R}_{i}$ is the location of vortex $i$ 
and $\eta$ is the damping coefficient.  
The repulsive vortex-vortex interaction force is
${\bf F}_{i}^{vv} = \sum^{N_{v}}_{j\ne i}f_{0}
K_{1}(R_{ij}/\lambda){\bf {\hat R}}_{ij}$, where
$K_{1}$ is the modified Bessel function,
$f_{0}=\phi_0^2/(2\pi\mu_0\lambda^3)$, 
$\phi_{0}=h/2e$ is the elementary flux quantum,
$R_{ij}=|{\bf R}_i-{\bf R}_j|$, and ${\bf \hat R}=({\bf R}_{i}-{\bf R}_j)/R_{ij}$.
The substrate force from $N_p$
parabolic traps of radius $r_p=0.35\lambda$ and maximum strength 
$F_{p}=1.85$ has the form 
${\bf F}^{vv}_i=\sum_{k=0}^{N_p}f_0 (F_p/r_p) \Theta(1-R_{ik}/R_p){\bf \hat R}_{ik}$,
where $\Theta$ is the Heaviside step function. 
At the matching field of $B_\phi$, $N_v/N_p=1$.
${\bf F}^{ext}_i$ is the driving force from an external current applied uniformly
to all particles.
We initialize the particle positions using simulated annealing, then slowly 
turn on an external drive
${\bf F}^{ext} = F_{D}(\cos\theta {\bf {\hat x}}+\sin\theta{\bf \hat y})$
with $\theta$ reported in degrees. 
We measure the velocity in the $y$-direction 
$V_y=\sum_{i=0}^{N_v}{\bf v}_i \cdot {\bf {\hat y}}$
as we increment the drive angle $\theta$.
All forces and lengths are measured in units of $f_{0}$ and $\lambda$.  

In Fig.~1(a) we plot the pinning site locations showing the 5-fold 
Penrose tiling and in 
Fig.~1(b) we show the positions of the non-driven pinned particle state
at 
$B/B_{\phi} = 1.61$. 
The structure factor $S(k)$
of the particle positions in this sample plotted in 
the inset of Fig.~1(d)  
has a 10-fold peak structure, indicative of quasicrystalline order 
such as that found in the strong substrate 
limit for colloidal systems \cite{Bechinger}. 
In general, for this $F_{p}$ the pinned states 
at different filling fractions have quasicrystalline ordering. 

\begin{figure}
\includegraphics[width=3.5in]{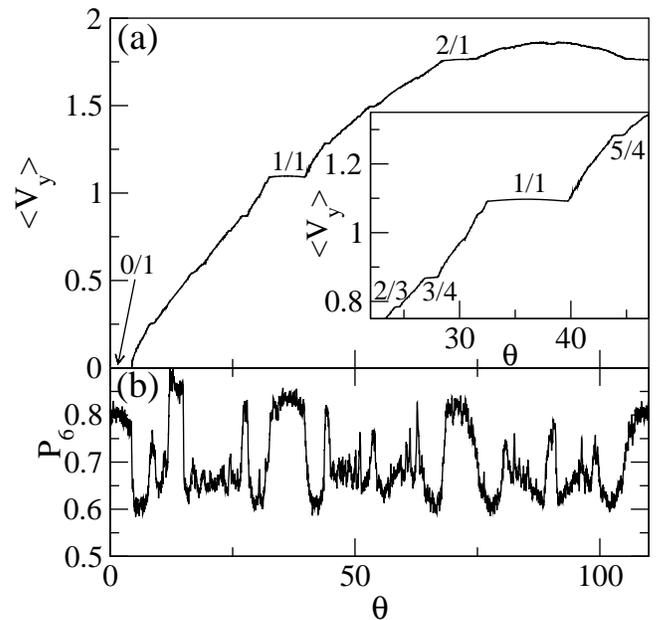}
\caption{
(a) The average velocity in the $y$-direction $\langle V_{y}\rangle$ 
vs drive angle $\theta$ 
for the system in Fig.~1(a) with $B/B_\phi=3.225$ at $F_D=2.0$.
A series of pronounced steps appear at multiples of $36^{\circ}$, 
marked 0/1, 1/1, and 2/1, when the particle motion becomes locked to
certain symmetry directions of the substrate.  
Locking at fractional multiples of $36^\circ$ produces
many smaller steps.  Inset: A blow up of the main panel 
near the 1/1 step showing some
of the fractional steps.
(b) The corresponding 
$P_6$ vs $\theta$ showing that the system is more ordered
on the locking steps.
}
\end{figure}

Figure~2(a) shows $\langle V_{y}\rangle$ versus the drive angle
$\theta$ for 
a sample with $B/B_\phi=3.225$ and
$F_{D} = 2.0$.  At this drive, all the particles
are in motion.   The clear set of steps in $\langle V_{y}\rangle$ 
is a signature of the directional locking that occurs
when the particles move along a single direction 
over a range of drive angles 
\cite{ReichhardtA,Ahn,Korda,Lacasta,Drazer}. 
For the same set of parameters but with randomly placed pinning, 
no steps occur in $\langle V_{y}\rangle$ \cite{Sn}.
The locking angles are related to the tenfold orientational ordering,
and occur at multiples of
$\theta=360^{\circ}/10 = 36^{\circ}$, as
highlighted in Fig.~2(a) for $\theta = 0^{\circ}$
(the 0/1 step), $36^{\circ}$ (the 1/1 step), and $72^{\circ}$ (the 2/1 step).
There are numerous smaller locking steps associated with rational fractions
of $36^{\circ}$, including
$1/4$, 1/2, 2/3, 3/4, 5/4, and $5/2$.  
Some of these smaller steps are visible in the inset of Fig.~2(a), which shows
a blowup of the main panel.
This result proves that it is possible
for directional locking to occur on quasiperiodic substrates, 
indicating that {\it orientational order} rather than translational order 
of the substrate
is the essential ingredient for directional locking. 
This opens the possibility of creating novel fractionation 
devices using quasiperiodic substrates, and may also be
relevant to understanding frictional studies 
performed with quasicrystalline substrates. 
We find that on the locking steps, there is a higher amount of order
in the particle lattice.
In Fig.~2(b) we plot the fraction of sixfold coordinated
particles $P_{6}=\sum_i^{N_v}\delta(z_i-6)$ as a function of $\theta$, 
where the particle coordination number $z_i$ is 
obtained from a Voronoi construction that is
not sensitive to square orderings.
In the pinned state $P_{6} = 0.525$, while for the
driven systems $P_6$ reaches its highest values on the locking steps
and is lower in the non-locking ranges of $\theta$.

\begin{figure}
\includegraphics[width=3.5in]{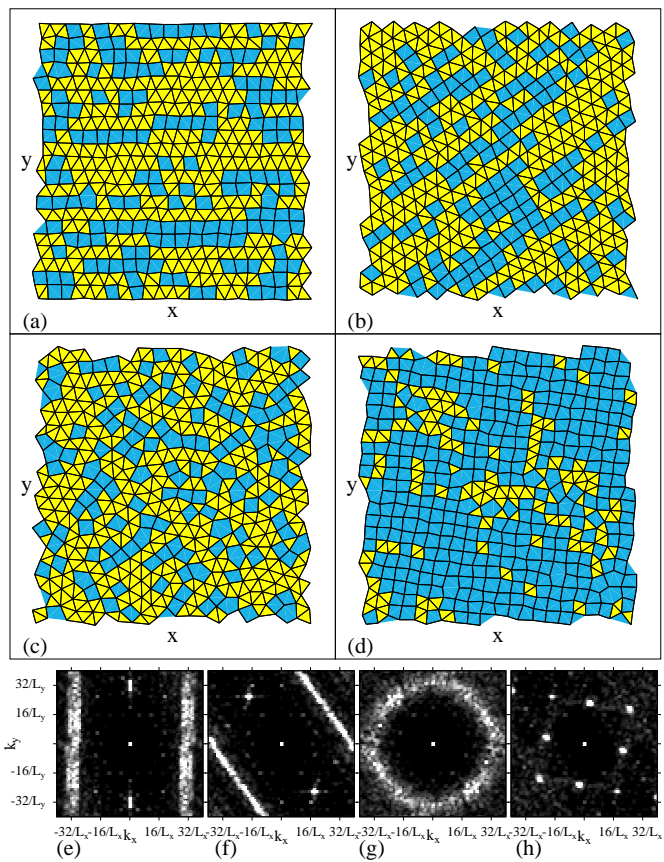}
\caption{
Delaunay triangulations of the particle 
positions in the sliding state. 
Bonds longer than $1.1a_0$ are omitted, where $a_0$ is the lattice spacing
of a triangular lattice with the same density.
Yellow (light) polygons are triangular and blue (dark) polygons are square
or nontriangular.
For a sample with $B/B_\phi=3.225$
and $F_D=2.0$ we show
(a) the 0/1 step at $\theta=0$ exhibiting the oriented triangles and squares
of the dynamically induced Archimedean tiling,
(b) the 1/1 step at $\theta=32^{\circ}$ where the Archimedean ordering 
aligns with the direction of motion,
and (c) $\theta=41^{\circ}$ in the unlocked region just above the 1/1 step. 
(d) The triangulation for the
1/1 step at $\theta=32^{\circ}$ for a sample with $B/B_\phi=1.61$
and $F_D=2.0$, where most of the tiles are square.
Smectic-like ordering appears in $S(k)$ corresponding to the 
Archimedean tilings of (e) panel (a) and
(f) panel (b).
(g) $S(k)$ for the unlocked state shown in panel (c) 
has a ring structure indicating liquid ordering.
(h) At the lower field of $B/B_\phi=1.61$,
$S(k)$ corresponding to panel (d) has square ordering.  
}
\end{figure}

We tessellate the positions of the particles in the moving state 
as in Refs.~\cite{Bechinger,C} in order to
reveal the nature of the dynamical ordering.
In Fig.~3(a) we show the tessellation of the moving state on the 
$0/1$ step from Fig.~2(a), where the flow is
locked in the $x$-direction.  
Both square and triangular tiles are present, and
the square tiles are aligned with the flow along the $x$ direction.
This state is very similar to the Archimedean type tiling 
found for colloids pinned by substrates of intermediate strength 
\cite{Bechinger,C}.  The smectic features of the corresponding structure
factor $S(k)$ in Fig.~3(e) indicate that unlike the static colloidal tiling,
the dynamical tiling does not have a one-dimensional (1D) quasiperiodic
structure, which would appear as a series of peaks
in $S(k)$ \cite{C}.
More recent colloid experiments reveal the development of
smeared smectic-like 
peaks in $S(k)$ similar to those in Fig.~3(e) when the strength
of the laser-induced substrate is increased \cite{G}.   
Moving smectic states also appear in systems with random pinning, but 
these states have a purely triangular tiling containing some dislocations 
\cite{Koshelev,Giamarchi}.

We find that oriented Archimedean tilings appear on each locking step, as 
illustrated for the 1/1 step in Fig.~3(b) and (f).
The particles move along 1D channels on the locked steps \cite{Sn}, but
follow winding and rapidly changing trajectories when not on the steps.
In Fig.~3(c) we show the tessellation for driving just above the 1/1 step in
a disordered flow regime.  The orientational ordering is lost and the
corresponding $S(k)$ in Fig.~3(g) has a ring structure characteristic of a
moving liquid \cite{Koshelev}. 
We find that the relative number of square and rectangular tiles appearing
on the locking steps varies as a function of filling.
For example, at $B/B_{\phi} = 1.61$ on the $1/1$ locking step,
Fig.~3(d) shows that the tiling is predominantly square, and
$S(k)$ in Fig.~3(h) has square ordering.

\begin{figure}
\includegraphics[width=3.5in]{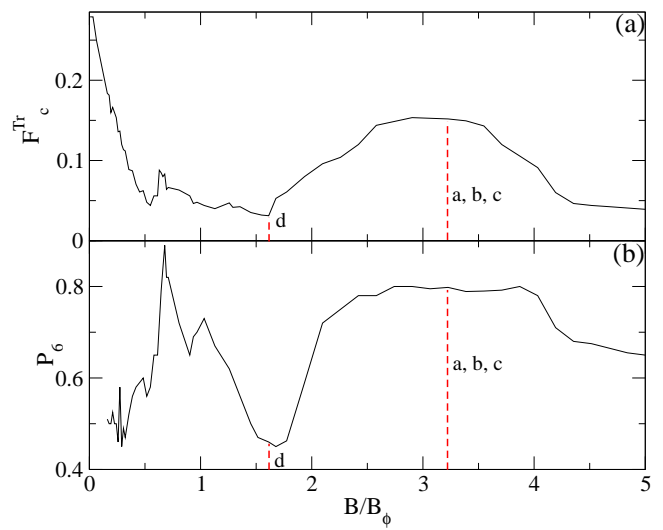}
\caption{ 
(a) The depinning threshold $F^{Tr}_{c}$ indicating the width of the 0/1 locking 
step vs $B/B_{\phi}$. 
(b) 
$P_{6}$ 
vs $B/B_{\phi}$. 
The broad peak in both quantities for $1.7 < B/B_{\phi} < 4.3$ 
appears when the system forms the dynamical Archimedean 
tiling illustrated in Fig.~3(a), while 
the dip at $B/B_{\phi} = 1.6$ corresponds to the moving square
structure shown in Fig.~3(h). 
Additional peaks in $P_6$ occur at $B/B_\phi=1.0$ and $B/B_\phi=1/\tau$,
where $\tau$ is the golden mean. 
Dashed lines and the labels a-d indicate the fields shown in Fig.~3(a-d).
}
\end{figure}

To understand where the different moving phases 
occur as well as the changes in the effectiveness of the directional locking, 
in Fig.~4(a) we plot the transverse depinning force $F^{Tr}_{c}$ determined
by $F^{Tr}_c=F_D\sin\theta_e$, where $\theta_e$ is
the end of the 0/1 locking step, versus $B/B_\phi$
and in Fig.~4(b) we show the corresponding 
$P_{6}$ versus $B/B_\phi$.
There is a broad maximum in $F^{Tr}_{c}$ 
associated with the Archimedean type ordering 
for $1.7 <B/B_{\phi} < 4.3$ 
and another sharper peak 
in $F^{Tr}_c$ at $B/B_{\phi} = 0.62$.  Both of these features 
are accompanied by peaks in $P_{6}$. 
Our transverse depinning measurements, in which the vortices
are depinning from a moving state, differ in several ways from the
longitudinal depinning from a static state
previously measured in simulations and
experiments \cite{MiskoKemler,Sihr,Mosh}. 
For example, $F^{Tr}_c$ shows no peaks
at $B/B_{\phi} =  0.81$ or $1.0$, although we do find a weak
peak in in $P_{6}$ at $B/B_{\phi} = 1.0$. 
In agreement with the experiment of Ref.~\cite{Sihr}, 
we observe a prominent peak in $F^{Tr}_c$ at $B/B_\phi=0.62=1/\tau$ where 
$\tau=(1+\sqrt{5})/2$ is
the golden mean.
The lowest value of $F^{Tr}_{c}$ occurs at $B/B_{\phi}= 1.61\approx \tau$. 
At this field, the widths of the locking steps 
are suppressed and the system has the square
ordering shown in Fig.~3(d,h). 
For $B/B_{\phi} > 4.0$, the ordering of the locked phases is reduced and
$S(k)$ develops a ring structure on the locking steps,
while the vortex-vortex interactions are strong enough in the nonlocking
regimes to produce sixfold ordering.
For $0.5 < B/B_{\phi} < 1.5$ the system generally exhibits
a smectic ordering containing few square tiles 
corresponding to a highly defected triangular lattice similar to that found
for driving over random substrates.  

We find that the ordered moving phases 
only occur when $F_D$ is sufficiently large.
For $F_D<1.25$ with $F_p=1.85$, significant plastic flow occurs,
$S(k)$ is always liquidlike, and the step structures in the velocity curves
disappear.
For $F_{D} \gg F_{p}$ the width of the steps slowly decreases with
increasing $F_D$.  We also find the same results reported here
for driven colloids interacting with a screened Coulomb 
interaction and a quasiperiodic substrate \cite{Sn}.  
Sevenfold or tetradecagonal quasiperiodic substrates also produce
directional locking, but the locking steps are significantly reduced
in width \cite{Sn}.

We have shown that interacting particles such as vortices or 
colloids driven over a quasiperiodic substrate 
exhibit rich dynamical behaviors including pronounced 
directional locking where the particles prefer to move 
at particular angles with respect to the substrate.
Directional locking has already been studied in periodic systems, but this
study is the first demonstration of directional locking on
quasiperiodic substrates.
For strong substrates, the pinned state has quasicrystalline 
ordering, but the moving state can have 
square or smectic ordering depending on the orientation of the drive.
At certain filling fractions the system 
forms a moving Archimedean tiling ordered state
similar to the pinned Archimedean tiling ordering observed
for colloids on decagonal substrates, 
but with smectic rather than quasiperiodic character.
We also show that the dynamically ordered states
produce distinct signatures in the
transverse depinning threshold.

This work was carried out under the auspices of the 
NNSA of the 
U.S. DoE
at 
LANL
under Contract No.
DE-AC52-06NA25396.

\section{SUPPLEMENTARY MATERIAL}

\vskip2pc
\subsection{Directional Locking for Vortices on Tetradecagonal Quasicrystalline Substrates}
Here, we show that the directional locking effects 
described in the main text for driven vortices moving 
over decagonal quasicrystalline substrates also occur 
when the vortices are driven over tetradecagonal or sevenfold-symmetric
quasicrystalline substrates such as that shown in Suppl. Fig.~1. 
We use the same pinning parameters as for the decagonal 
substrate with $F_{p} = 1.85$ and $r_{p} = 0.35\lambda$, and apply a drive
$F_D=2.0$ at $B/B_{\phi} = 3.9$. 
In Suppl. Fig.~2(a) we plot $\langle V_{y}\rangle$ versus 
the drive angle $\theta$ and in Suppl. Fig.~2(b) we show
the fraction of sixfold coordinated particles $P_{6}$ versus $\theta$. 
There is a clear set of steps in $\langle V_y\rangle$ which are
accompanied by enhanced sixfold ordering as shown by the increases in
$P_6$.
For the tetradecagonal substrate potential, 
the major locking steps occur at drive angles that are
integer multiples of $360^\circ/14$, producing the
1/1 step at $25.7^\circ$, the 2/1 step at $51.42^\circ$, 
and the 3/1 step at $77^\circ$.  For the
decagonal substrate the locking steps fall at integer multiples 
of $360^\circ/10$, as described in the main text. 
Fractional locking steps are nearly absent for the tetradecagonal substrate;
however, for lower fillings it is possible to resolve some
fractional steps. 
This is shown in Suppl. Fig.~3(a,b) for $B/B_{\phi} = 2.96$.  The
fractional steps $m/4$ with $m$ integer are the most pronounced and
are associated with enhanced sixfold ordering in $P_{6}$. 
This result indicates that directional locking is a generic 
feature for particles moving over different types of quasicrystalline 
substrates. The width of the locking steps is generally much narrower for
the tetradecagonal substrate than for the decagonal substrate.
Additionally, although the decagonal substrate
produced pronounced five-fold ordering of the particles for some
fillings, we do not find
strong seven-fold particle ordering on the tetradecagonal substrate.
Instead, on the locking steps, we observe
smectic ordering composed of sixfold-coordinated
particles and a limited number of dislocations.

\setcounter{figure}{0}
\begin{figure}
\includegraphics[width=3.2in]{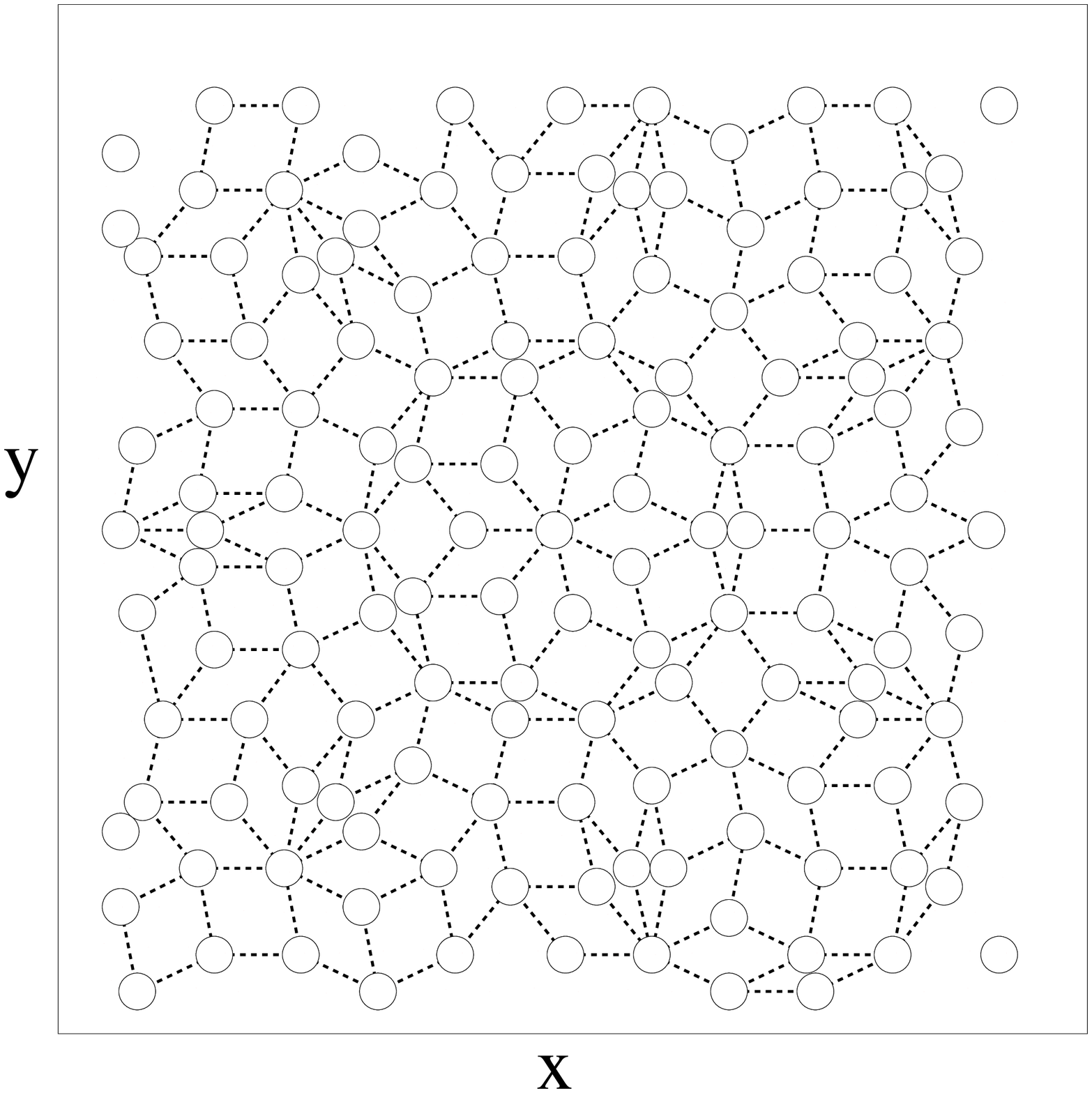}
\caption{
The locations of the pinning sites (open circles) for a portion of the
sample with seven-fold or tetradecagonal ordering.  Dashed lines indicate the
positions of the tiles used to define the pinning locations.
}
\end{figure}

\begin{figure}
\includegraphics[width=3.2in]{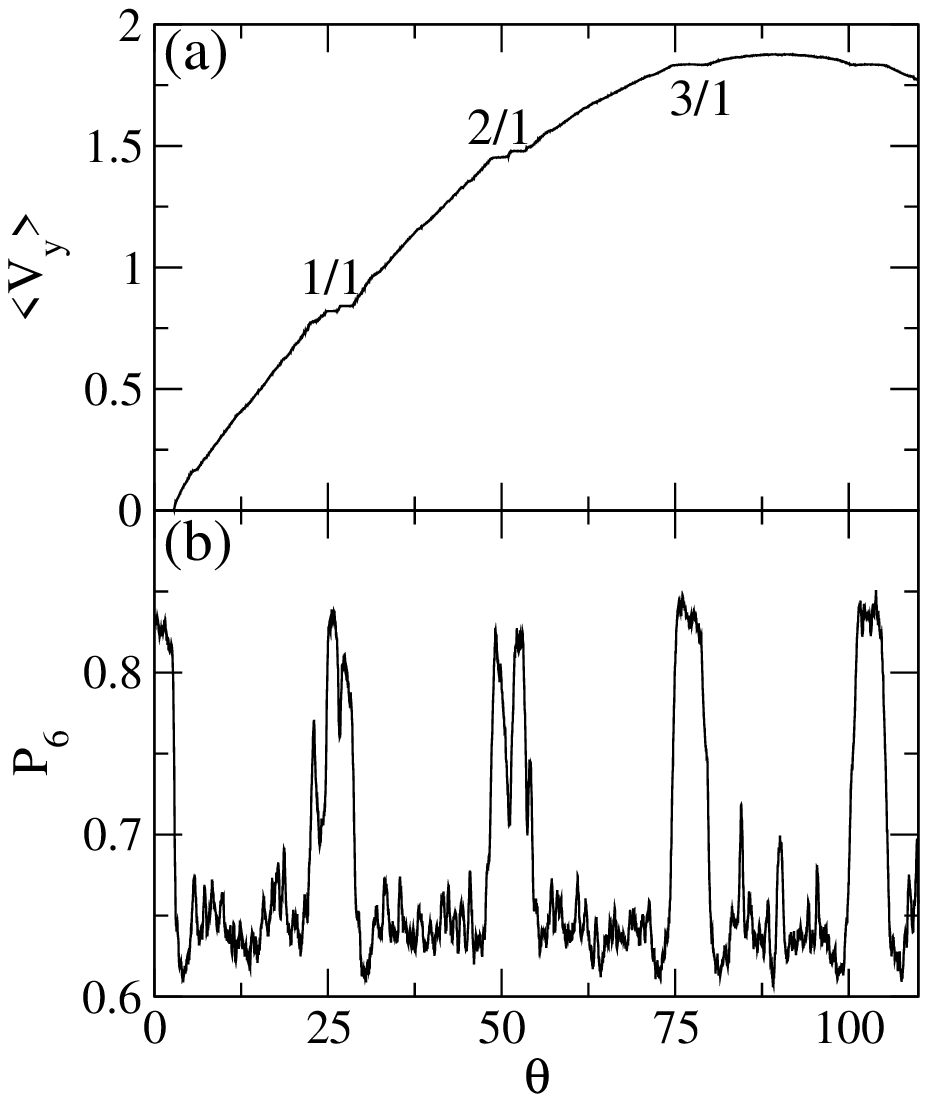}
\caption{
(a) The average velocity in the $y$-direction $\langle V_{y}\rangle$ 
vs drive angle $\theta$ for the system in Suppl. Fig.~1 
with a sevenfold quasicrystalline pinning array for 
$F_{p} = 1.85$, $r_{p} = 0.35\lambda$, $B/B_{\phi} = 3.9$, and $F_D=0.2$.
Several steps appear at the directional locking angles which are
integer multiples of $360^\circ/14$.
The 1/1, 2/1, 3/1, and 4/1 lockings are clearly visible.
(b) The corresponding $P_6$ vs $\theta$
shows that on the locking steps the system develops a considerable 
amount of sixfold ordering.   
}
\end{figure}

\begin{figure}
\includegraphics[width=3.2in]{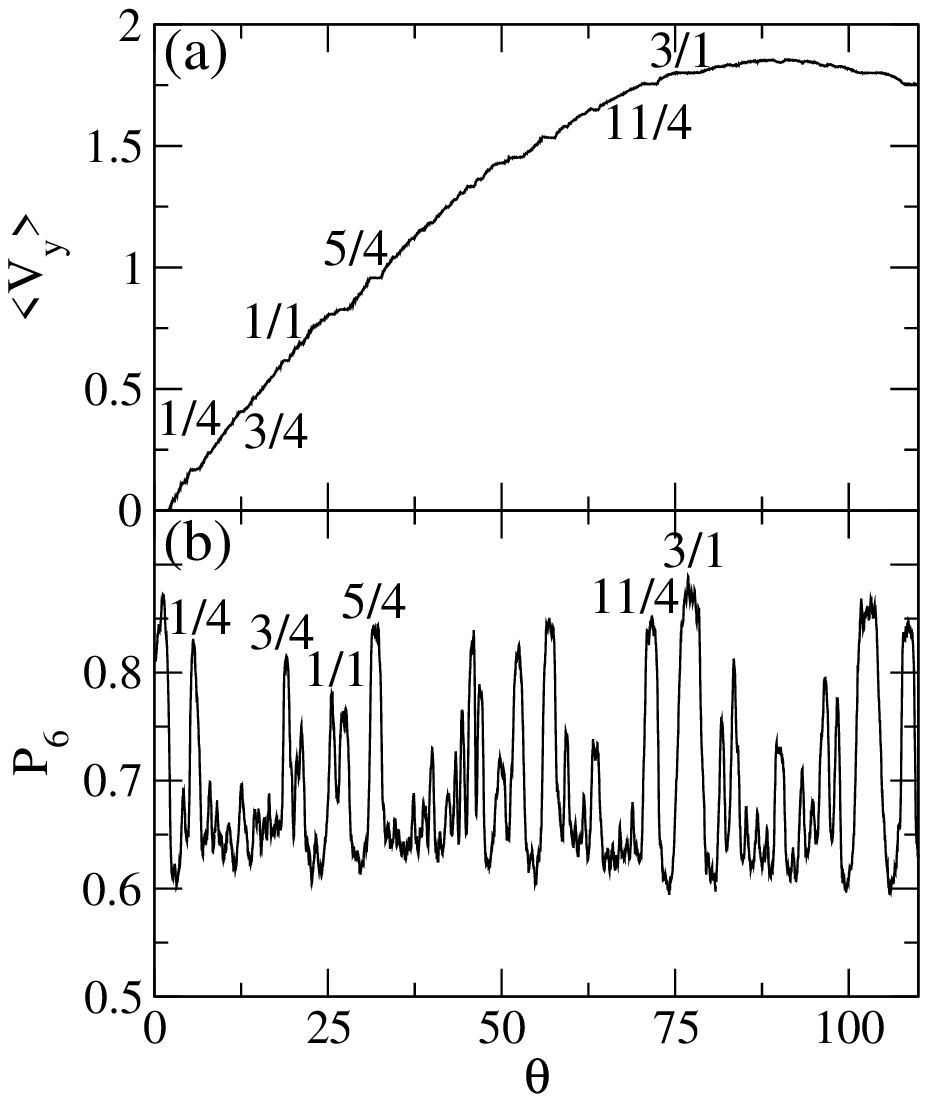}
\caption{
(a) $\langle V_y\rangle$ vs drive angle $\theta$ for the system in Suppl. Fig.~2
but for a filling of $B/B_{\phi} = 2.96$ where additional fractional locking
steps are visible, particularly at fractions of $m/4$ where $m$ is an
integer.  
Some representative steps are marked.
(b) The corresponding $P_6$ vs $\theta$.
}
\end{figure}

\subsection{Directional Locking for Colloids on Decagonal and Tetradecagonal 
Substrates} 

We next show that the same directional locking effects 
described in the main text for vortices driven over 
quasicrystalline substrates also occurs for colloidal particles
driven over quasicrystalline substrates.  
We model a two-dimensional system of $N_c$
interacting colloids in the presence of fivefold and sevenfold 
quasicrystalline substrates.
The substrate is composed of localized pinning traps 
with maximum strength $F_{p}$ and radius $r_{p}=0.35$. 
We simulate the motion of the colloids 
using the same procedure used previously to model colloid 
dynamics on random substrates \cite{Ref1} and periodic substrates \cite{Ref2}, 
by integrating the following equation of motion: 
$\eta d{\bf R}_{i}/dt = {\bf F}^{cc}_{i} + {\bf F}^{s}_{i} + {\bf F}_{D}$.
Here ${\bf R}_{i}$ is the location of colloid $i$ 
and $\eta$ is the damping coefficient.  
The colloid-colloid interaction potential has the Yukawa form
$V(R_{ij}) = (E_{0}/R_{ij})\exp(-\kappa R_{ij})$, where 
$R_{ij} = |{\bf R}_{i} - {\bf R}_{j}|$, 
$E_{0} = Z^{*2}/(4\pi\epsilon\epsilon_{0})$, $\epsilon$ is the
solvent dielectric constant, $Z^{*}$ is the effective charge of each colloid, 
and $1/\kappa$ is the screening length.  
The colloid-colloid interactions are repulsive and are given by
${\bf F}^{cc}_{i} = -\sum^{N_{c}}_{j\neq i}\nabla V(R_{ij})$. 
Lengths are measured in units of $a_{0}$ and forces in units of 
$F_{0} = E_{0}/a_{0}$. 
The substrate force term ${\bf F}_i^s$ 
comes from $N_{p}$ pinning sites placed in a decagonal or tetradecagonal pattern 
and has the same form as described for the vortex system. 
The colloid density relative to the pinning density is $N_c/N_p$.
We neglect hydrodynamic interactions in the colloidal system and assume that
the colloids are in the low volume fraction, highly charged, 
electrophoretically driven limit. 
The external force ${\bf F}_D$ is the same as that used for the vortex case
and 
we measure $\langle V_{y}\rangle$ and $P_{6}$ as a function of
the drive angle $\theta$. 

\begin{figure}
\includegraphics[width=3.2in]{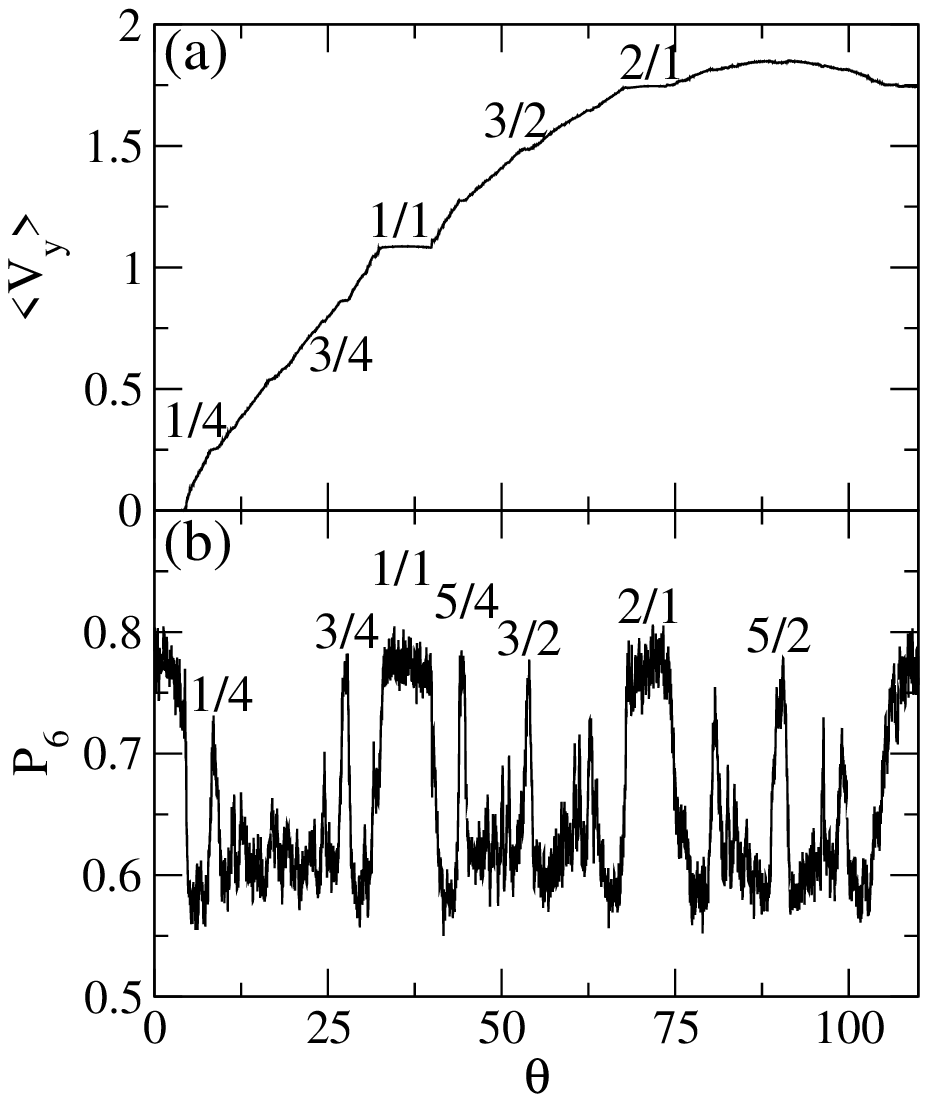}
\caption{ 
The directional locking for charged colloidal particles 
driven over a decagonal substrate
in the same manner as the vortices in the main text.
Here $N_{c}/N_{p} = 2.9$ and $F_{p} = 0.75$.
In general, the same features observed for the vortex system 
also appear for the colloidal system. 
(a)$ \langle V_{y}\rangle$ vs $\theta$ 
with the main locking steps at integer multiples of $360^{\circ}/10$ 
highlighted at 1/1 and 2/1.  Several fractional steps 
also appear at rational fractions of the integer steps  
such as at $1/4$, $3/4$, and $3/2$. 
(b) The corresponding $P_{6}$ vs $\theta$ shows that along 
the steps the system has higher ordering.
On the main integer matching steps, the system 
forms the dynamically induced Archimedean ordering also observed
in the vortex system (as described in the main text).
}
\end{figure}

In Suppl.~Fig.~4(a) we plot $\langle V_{y}\rangle$ versus $\theta$ 
for a colloidal system on a decagonal substrate with $F_{p} = 0.75$ 
at $N_{c}/N_{p} = 2.9$,
and in Fig.~4(b) we show the corresponding $P_{6}$ versus $\theta$.
Here, all the general features of the directional locking 
in the vortex system also appear
for the colloidal system, 
including the dominant locking steps at 
integer multiples of $360^{\circ}/10$ such as
1/1 and 2/1 and additional
fractional lockings at 1/2, 3/2, 5/2, 1/4, and 3/4.
Each locking step is associated with enhanced ordering of the 
colloidal lattice when the system forms a dynamically ordered Archimedean
tiling state. 

\begin{figure}
\includegraphics[width=3.2in]{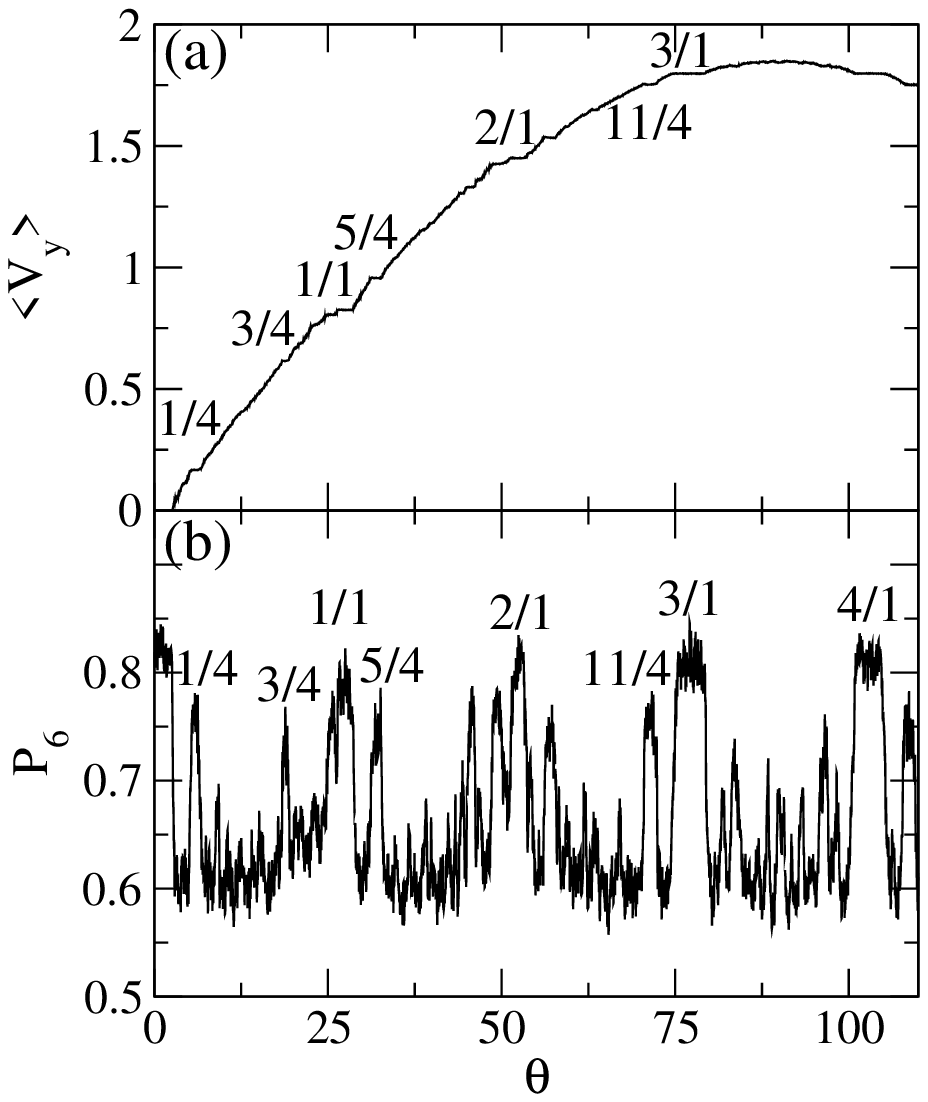}
\caption{ 
Directional locking for charged colloidal particles driven 
over a sevenfold or tetradecagonal substrate for the same parameters 
in Suppl.~Fig.~4.
(a) $\langle V_y\rangle$ vs $\theta$ and (b) $P_6$ vs $\theta$. 
The same features found for the vortex system 
in Suppl.~Fig.~3 appear for the colloid system, including
locking steps falling at integer multiples of $360^{\circ}/14$. 
We also observe the strongest fractional locking steps at
fractions $m/4$ with $m$ integer.
}
\end{figure}

In Suppl.~Fig.~5(a,b) we show $\langle V_y\rangle$ and $P_6$ versus $\theta$
for colloids moving over a tetradecagonal substrate.
We find the same features illustrated in Suppl.~Fig.~3 for vortices
moving over a tetradecagonal substrate.
The dominant locking steps occur at integer multiples of $360^\circ/14$. 
In general, the locking steps are weaker than for the decagonal substrate
and the dominant fractional steps occur at ratios of $m/4$
with $m$ integer.  The fractional steps are also associated with
a series of peaks in $P_{6}$, where additional fractions beyond $m/4$ appear
as smaller peaks that do not show full ordering.
These results indicate that the directional locking effects on 
quasicrystalline periodic substrates can
be observed for various types of interacting particles
including colloids and vortices, and 
that the directional locking is a robust feature in these systems.

\end{document}